\documentclass[aps,preprint,showpacs,amsmath,amssymb]{revtex4}

\usepackage{graphicx}
\usepackage{dcolumn}
\usepackage{bm}

\begin{document}

\preprint{Bailly RScI 2009}

\title{Highly sensitive frequency metrology for optical anisotropy measurements}

\author{Gilles Bailly}
\author{Rapha\"{e}l Thon}
\author{C\'ecile Robilliard}
\email{cecile.robilliard@irsamc.ups-tlse.fr}

\affiliation{Universit\'e de Toulouse-UPS; Laboratoire Collisions Agr\'egats R\'eactivit\'e, IRSAMC; F-31062 Toulouse, France. \\
CNRS, UMR 5589; F-31062 Toulouse, France.}

\date{\today}

\begin{abstract}
In this paper we present a novel apparatus aimed at measuring very small birefringences and anisotropies, based on frequency metrology and not on polarimetry as usual. In our experiment, a very high finesse resonant cavity is used to convert the phase difference into a resonance frequency difference, which can then be measured with very high accuracy. We describe the set-up and present the results of experimental tests which exhibited a sensitivity $\delta n \simeq 2\times10^{-18}$, allowing for the measurement of long-predicted magneto-electro-optical effects in gases. Since the shot-noise limited sensitivity of our apparatus lies well below the state-of-the-art sensitivity, frequency metrology appears as a promising technique for small birefringence measurements. 
\end{abstract}

\pacs{33.57.+c, 42.25.Lc, 42.62.Eh, 06.30.Ft}
\maketitle

\section{Introduction \label{introduction}}

High finesse optical cavities play an important role in modern physics, especially in precision experiments: one famous example concerns the gravitational wave searches \cite{LIGO,VIRGO,GEO600,TAMA,AIGO} where relative sensitivities range from $\gamma_{\mathrm{rel}} \sim 10^{-21}/\sqrt{\mbox{Hz}}$ to $\gamma_{\mathrm{rel}} \sim 2\times 10^{-23}/\sqrt{\mbox{Hz}}$, but many other experiments include high finesse cavities, such as optomechanics measurements \cite{Arcizet2008} achieving $\gamma_{\mathrm{rel}} \sim 2\times 10^{-17}/\sqrt{\mbox{Hz}}$ or tests of the isotropy of light propagation \cite{Eisele2009} where $\gamma_{\mathrm{rel}} \sim 1\times 10^{-17}/\sqrt{\mbox{Hz}}$.

Since the first electro- and magneto-optical effects have been discovered, about 150 years ago, the sensitivity of the experiments has improved by more than 10 orders of magnitude. In particular, the insertion of an optical cavity -- resonant or multipass -- between the polarizer and the analyzer, first suggested by E.~Zavattini as early as 1979 \cite{Zavattini1979}, brought the sensitivity down by several orders of magnitude thanks to the accumulation of the birefringence effect on a longer optical path. This approach has been widely used for the past 20 years, leading in particular to several measurements of Kerr \cite{Read1997,Inbar2000,Bielsa2005,Durand2009} and Cotton-Mouton \cite{Cameron1991,Brandi1998,Bregant2009,Mei2009} effects in gases. 

Table \ref{tab:sensitivity} presents an overview of the most recent experimental results. We use two quantities to compare the achieved sensitivities: the sensitivity $\gamma_n$, {\it i.e.} the noise equivalent birefringence at the relevant detection frequency, and the value $\delta n$ of the smallest birefringence actually measured, as inferred from the published data. The former is very useful to compare the sensitivity of each experiment to the shot-noise limit, but since the measurement times range from a few tens of ms \cite{Bielsa2009} to several tens of ks \cite{Zavattini2008,Muroo2003}, it cannot be used to directly compare the actual sensitivities of different experimental set-ups. The relevant parameter for this purpose is the smallest measured birefringence: it ranges from $1\times 10^{-13}$ to $1\times 10^{-17}$ for the experiments using the Pound-Drever-Hall frequency stabilization scheme, which is to be compared with the shot-noise limited birefringence $2\times 10^{-15}$ measured without any cavity at Tokyo University \cite{Muroo2003}. The present smallest value $4\times 10^{-19}$ was measured with an original locking scheme using optical feedback \cite{Durand2009}; note that it is less than one order of magnitude above the shot-noise limit. All these results were obtained after several years of development. 

\begin{table*}[!htb]
\label{tab:sensitivity}
\begin{tabular}{|c|c|c|c|c|c|}
\hline
Year & Group & Effect & Technique & $\gamma_n\; (1/\sqrt{\mathrm{Hz}})$ & $\delta n$ \\
\hline \hline
2000 & Tel-Aviv \cite{Inbar2000} & Kerr & Resonant cavity ellipsometer & $3\times 10^{-15}$ & $1\times 10^{-13}$ \\
\hline
2005 & BMV \cite{Bielsa2005} & Kerr & Resonant cavity ellipsometer & $2\times 10^{-15}$ & $1\times 10^{-15}$ \\
\hline
2009 & PVLAS \cite{Zavattini2008,Bregant2009} & Cotton-Mouton & Resonant cavity ellipsometer & $5\times 10^{-17}$ & $1\times 10^{-17}$ \\
\hline
2009 & Q \& A \cite{Mei2009} & Cotton-Mouton & Resonant cavity ellipsometer & $1\times 10^{-15}$ & $1\times 10^{-15}$ \\
\hline
2009 & BMV \cite{Battesti2008,Bielsa2009} & Cotton-Mouton & Resonant cavity ellipsometer & $2\times 10^{-18}$ & $2\times 10^{-15}$ \\
\hline
2009 & LASIM/LSP \cite{Durand2009} & Kerr & Resonant cavity ellipsometer & $8\times 10^{-19}$ & $4\times 10^{-19}$ \\
\hline \hline
2003 & Tokyo \cite{Muroo2003} & Cotton-Mouton & Single pass ellipsometer & $3\times 10^{-13}$ & $2\times 10^{-15}$ \\
\hline \hline
2000 & JILA \cite{Hall2000} & Sensitivity test & Frequency measurement & $2\times 10^{-15}$ & $1\times 10^{-17}$ \\
\hline
2009 & This work & Sensitivity test & Frequency measurement & $1\times 10^{-16}$ & $2\times 10^{-18}$ \\
\hline
\end{tabular}
\caption{Review of recent experimental results concerning small birefringence measurements in gases. The reported $\gamma_n$ is the noise equivalent birefringence at the detection frequency, while $\delta n$ corresponds to the smallest measured birefringence value inferred from the published data.}
\end{table*}

In 2000, an alternative approach was proposed by J.L.~Hall and coworkers \cite{Hall2000}, who suggested converting birefringence effects into frequency effects. This can be done by measuring the resonance frequency of a Fabry-Perot cavity. Indeed, this approach takes advantage of the recent progress in frequency metrology and should result in an unsurpassed sensitivity along with an impressive dynamic range. The tests performed by J.L.~Hall and coworkers lead to an equivalent sensitivity $\delta n = \delta \nu / \nu \simeq 1\times 10^{-17}$ (see Table \ref{tab:sensitivity}), which appeared to be limited by various experimental effects, including vibrations, perfectible optical isolation and birefringence noise of the cavity mirrors.

In fact, J.L.~Hall's proposal is similar to the method used in large active ring lasers aimed at precisely measuring the Earth rotation rate \cite{Schreiber2004,Hurst2009} thanks to the Sagnac effect. These rings have areas of tens to hundreds of square meters, and exhibit a relative frequency sensitivity on the order of $10^{-18}-10^{-20}/\sqrt{\mbox{Hz}}$.

In this paper, we present promising sensitivity tests performed on an apparatus which is based on the same idea but aims at measuring small directional anisotropies of the propagation medium. Indeed, we have built a ring cavity, so that one can distinguish two propagation directions. A laser light beam is frequency-stabilized to the clockwise cavity resonance, and injected into the cavity both clockwise and counterclockwise. The counterclockwise error signal yields the value of the directional anisotropy.

Our first goal is to measure the magneto-electric directional anisotropy of light which occurs in any medium in the presence of crossed transverse electric and magnetic fields. This anisotropy has indeed been predicted in the 70's \cite{Baranova1977} and has only been detected recently on light absorption by a crystal \cite{Roth2002}.

\section{Optical set-up \label{optics}}

The light source is a commercial NPRO Nd:YAG laser (Lightwave Electronics, now JDSU, NPRO-126) emitting $130\;$mW of cw light at $\lambda =1064\;$nm. It is tunable thanks to a Peltier cell controlling the crystal temperature and to a piezo-electric actuator that constrains the crystal.

\begin{figure}[!htb] 
\resizebox{1\columnwidth}{!}{\includegraphics{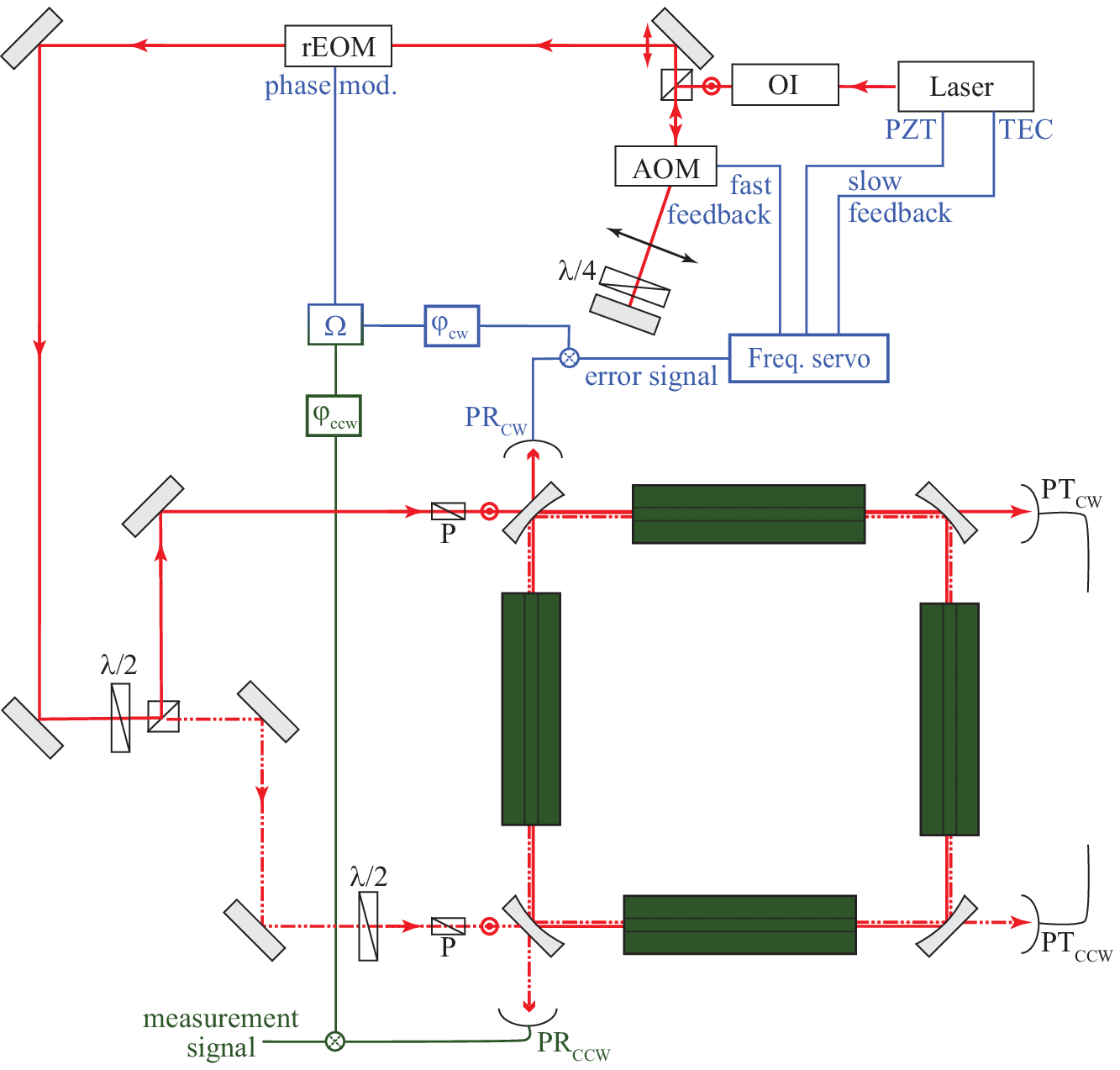}}
\caption{Experimental set-up (color online). The beampath is represented in red, the frequency stabilization system in blue and the measurement signal generation in green. The green blocks on each cavity arm represent magnetic and/or electric fields zones that can be inserted in order to measure magneto-electro-optic effects in gases. An optical isolator (OI) prevents feedback noise; the laser beam frequency is then frequency-shifted with an acousto-optic modulator (AOM) in a cat's eye retroreflector. A resonant electro-optic modulator (rEOM) provides the phase modulation at frequency $\Omega$ for the Pound-Drever-Hall frequency stabilization. The servo actuators are the laser thermo-electric cooler (TEC), the laser piezo-electric transducer (PZT) and the AOM. The light polarization is controlled all along the beampath by halfwave ($\lambda/2$) and quarterwave ($\lambda/4$) retardation plates, and by polarizers (P). Light is injected into the cavity both in the clockwise (cw) and counterclockwise (ccw) directions; the $\mathrm{PR_{cw}}$ and $\mathrm{PR_{ccw}}$ (resp. $\mathrm{PT_{cw}}$ and $\mathrm{PT_{ccw}}$) photodiodes monitor the reflected (resp. transmitted) power in both directions.\label{fig:setup}}
\end{figure}

\begin{figure}[!htb] 
\resizebox{0.6\columnwidth}{!}{\includegraphics{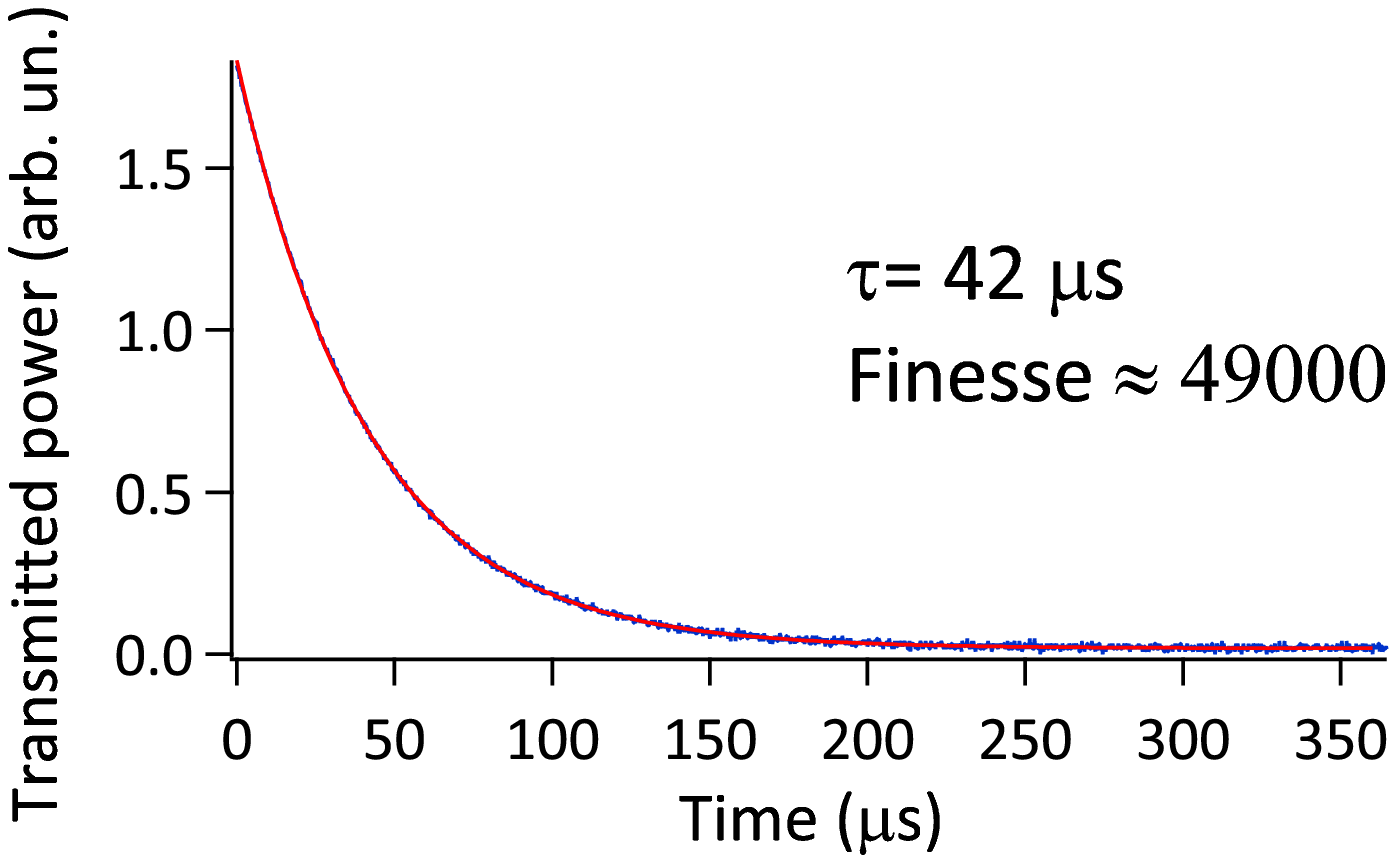}}
\caption{Typical finesse measurement. After the laser is rapidly switched off, the intracavity power decays exponentially with a time constant proportional to the finesse $\mathcal{F}$.\label{fig:finesse}}
\end{figure}

The core of the experimental set-up consists of a plane square optical cavity (see Fig. \ref{fig:setup}) made with 4 identical commercial mirrors from Layertec GmbH. These spherical mirrors with a $500\;$mm radius of curvature are arranged on a square with an arm length $L_0=400\;$mm, so that the beam waist on the mirrors is $w=0.7\;$mm. Due to non-normal incidence on the mirrors, a square cavity is strongly birefringent, so that in our case the resonance frequencies of the two polarization eigenmodes are well separated. We can thus adjust the incident polarization to be orthogonal to the cavity plane (s-polarization), so that only the high finesse mode is fed. The value of this finesse is $\mathcal{F}=\Delta \nu_{\mathrm{FSR}} / \Delta \nu_c\simeq 50000$, corresponding to a photon lifetime in the cavity $\tau = 42\; \mu$s (see Fig. \ref{fig:finesse}). Here $\Delta \nu_{\mathrm{FSR}} = 188\;$MHz is the free spectral range and $\Delta \nu_c$ is the cavity linewidth. 

The laser beam is matched to the cavity transverse mode TEM 00 by two pairs of cylindrical lenses, in order to take into account the astigmatism of the laser and that of the cavity. The position of the 4 lenses is adjusted by sweeping the laser frequency and minimizing the amplitude of higher order transverse modes. About $99\%$ of the intracavity power resonates in the TEM 00 mode. Light is injected into the cavity both clockwise (cw) and counterclockwise (ccw), and 4 large area homemade photodetectors monitor the light transmitted and reflected by the cavity in each direction. 

At this stage of the development, the cavity is not yet in a vacuum chamber but in an air-tight plexiglas box. A second wooden box backed with an acoustic isolating foam surrounds the first one to provide passive thermal insulation and airborne acoustic noise reduction. As far as solidborne noise is concerned, the whole experiment takes place on a Newport RS4000 optical table sustained by 4 PL2000 pneumatic isolators.

\section{Laser stabilization \label{stabilization}}

The laser is frequency stabilized on the cavity resonance for the clockwise laser beam, following the well-known Pound-Drever-Hall scheme \cite{Drever1983,Black2000}. 

The laser beam is frequency-shifted with an AOM mounted in a cat's eye retroreflector to circumvent the angular dependence of the diffracted beam on the RF frequency. Then it is injected clockwise (cw) into the square cavity. The reflected beam provides the measurement of the beam detuning with respect to the cavity, thanks to the rEOM phase modulation at frequency $\Omega =10\;$MHz (see Fig. \ref{fig:setup}). The modulation depth is set to $\beta \simeq 1$, so that the optical power $P_0$ is essentially concentrated in the carrier of power $P_c=J_0^2 (\beta) P_0$ and in the two first order sidebands $P_s=J_1^2 (\beta) P_0$. The reflected power is monitored with the PR$_{\mathrm{cw}}$ photodetector and its spectrum is shifted towards null frequency with a RF-phase sensitive demodulator which phase $\varphi_{\mathrm{cw}}$ is adjusted so as to maximize the signal amplitude accross a resonance.

This demodulated signal is null at resonance and linear for laser-cavity detunings smaller than the cavity linewidth, with a slope $D=4\sqrt{P_c P_s}/ \Delta \nu_c$.  It constitutes the frequency error signal which is fed to the frequency actuators through the servo-loop filters: a fast and fine feedback is provided by the AOM, while the laser piezo-transducer (PZT) and thermo-electric cooler (TEC) actuators account for slower and coarser corrections. The fast controller consists of 3 Proportional Integral (PI) stages with respective corner frequencies 30~kHz, 5~kHz and 3~kHz. The loop gain is adjusted to have a resonance of about 10~dB at the loop resonance frequency $f_{r} \simeq 180\;$kHz. Since the unity gain frequency is 2 to 3 times smaller for a gain margin between 6 and 10 dB, we can estimate the servo-loop bandwidth to 60-90 kHz. The fast controller output is sent on the one hand to the AOM driver, and on the other hand to the medium speed PZT-controller, which consists of a Proportional Integral Differential (PID) stage with an integrator (resp. differentiator) corner frequency $100\;$Hz (resp. $1\;$kHz). The proportional gain is set to damp smoothly the AOM command signal towards zero. A similar approach is adopted for the slow TEC-controller, with an integrator (resp. differentiator) corner frequency $15\;$mHz (resp. $150\;$mHz) nulling the mean value of the PZT command signal.

\begin{figure}[!htb] 
\resizebox{1\columnwidth}{!}{\includegraphics{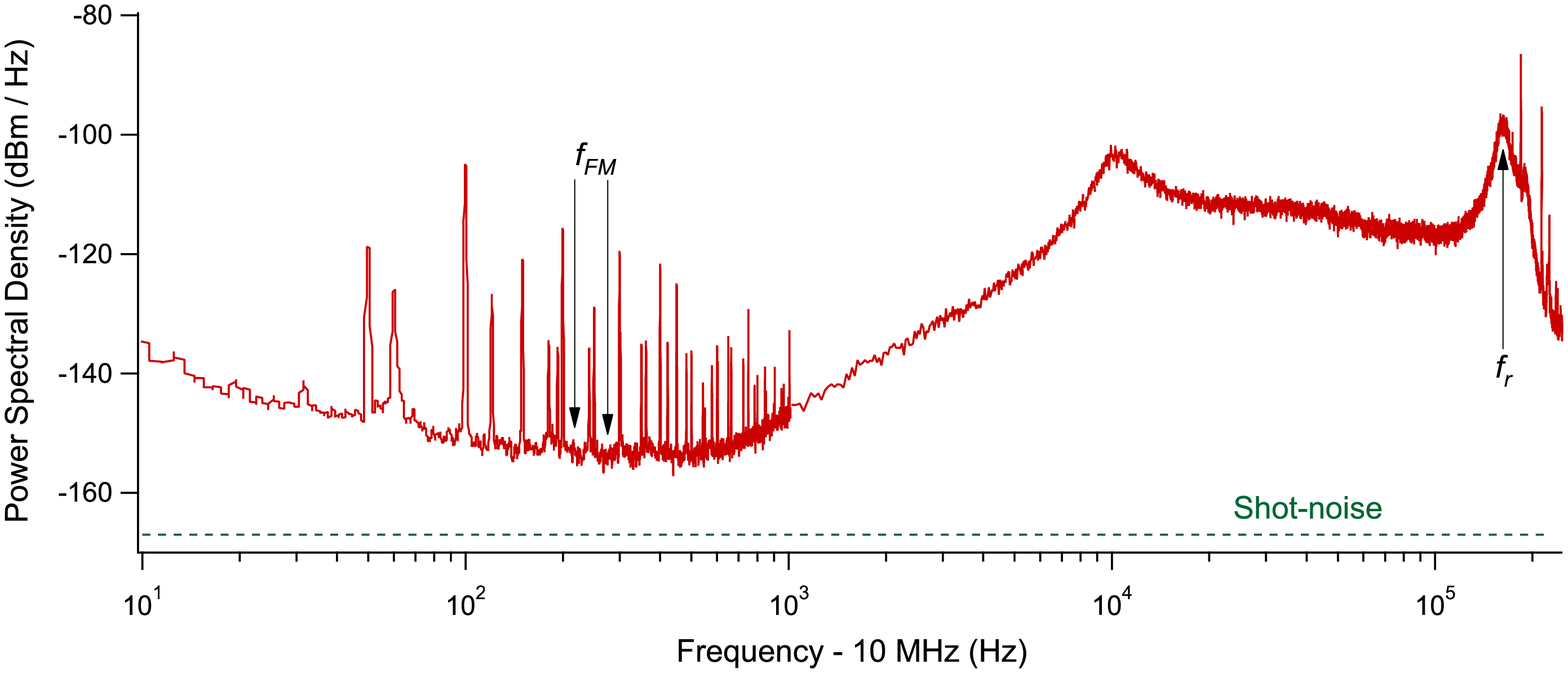}}
\caption{Spectrum analysis of the reflected signal before demodulation. In our present experimental conditions, it represents fairly well the frequency detuning noise. The frequency resolution is limited by our spectrum analyzer $1\;$Hz resolution bandwidth. The shot-noise limit represented as a dashed line on the figure is evaluated from the average power impinging on the photodiode PR$_{\mathrm{cw}}$. It is situated about 15~dB under the noise spectral density in the region of interest. The bounce at resonance frequency $f_r$ is clearly visible. The two values of $f_{\mathrm{FM}}$ used in our experiment are chosen so that the noise spectral density is at its lowest.\label{fig:spectrum}}
\end{figure}

To characterize the performance of the frequency stabilization, we intercept the signal immediately before demodulation via a 20~dB coupler. We present on Fig.~\ref{fig:spectrum} the power spectral density of this signal, obtained with a 3251 spectrum analyzer from Aeroflex. In our present experimental conditions, this spectrum represents fairly well the frequency detuning noise of the laser beam to the cavity. The smallest noise density is obtained at frequencies on the order of a few hundred Hz. The shot-noise limit can be evaluated from the average reflected power; it is situated about 15 dB below the lowest noise density.

We will see in the next section that the anisotropy signal appears as a disturbance of the laser frequency, hence the better noise suppression, the smaller anisotropy detected.

\section{Sensitivity measurements \label{sensitivity}}

In this section, we describe the experimental tests we have performed in order to evaluate the sensitivity of our apparatus. Since its present configuration aims at measuring frequency differences between the two propagation directions, we mimicked such an anisotropy with a broadband electro-optic modulator (bEOM, not shown on Fig. \ref{fig:setup}) placed on the clockwise beampath just before the cavity. The bEOM is fed with a sinusoidal signal of amplitude $V_{\mathrm{FM}}$ at frequency $f_{\mathrm{FM}}$, giving birth to a phase modulation of amplitude $\tilde{\beta} V_{\mathrm{FM}}$ at frequency $f_{\mathrm{FM}}$, hence a frequency modulation of amplitude $\delta \nu_{\mathrm{FM}} = \tilde{\beta} V_{\mathrm{FM}} f_{\mathrm{FM}}$ at the same frequency $f_{\mathrm{FM}}$. Here $\tilde{\beta} = 0.85\;$mrad/V is the bEOM modulation depth, which has been accurately measured at this wavelength in a separate experiment with a Michelson interferometer.

\begin{figure}[!htb] 
\resizebox{1\columnwidth}{!}{\includegraphics{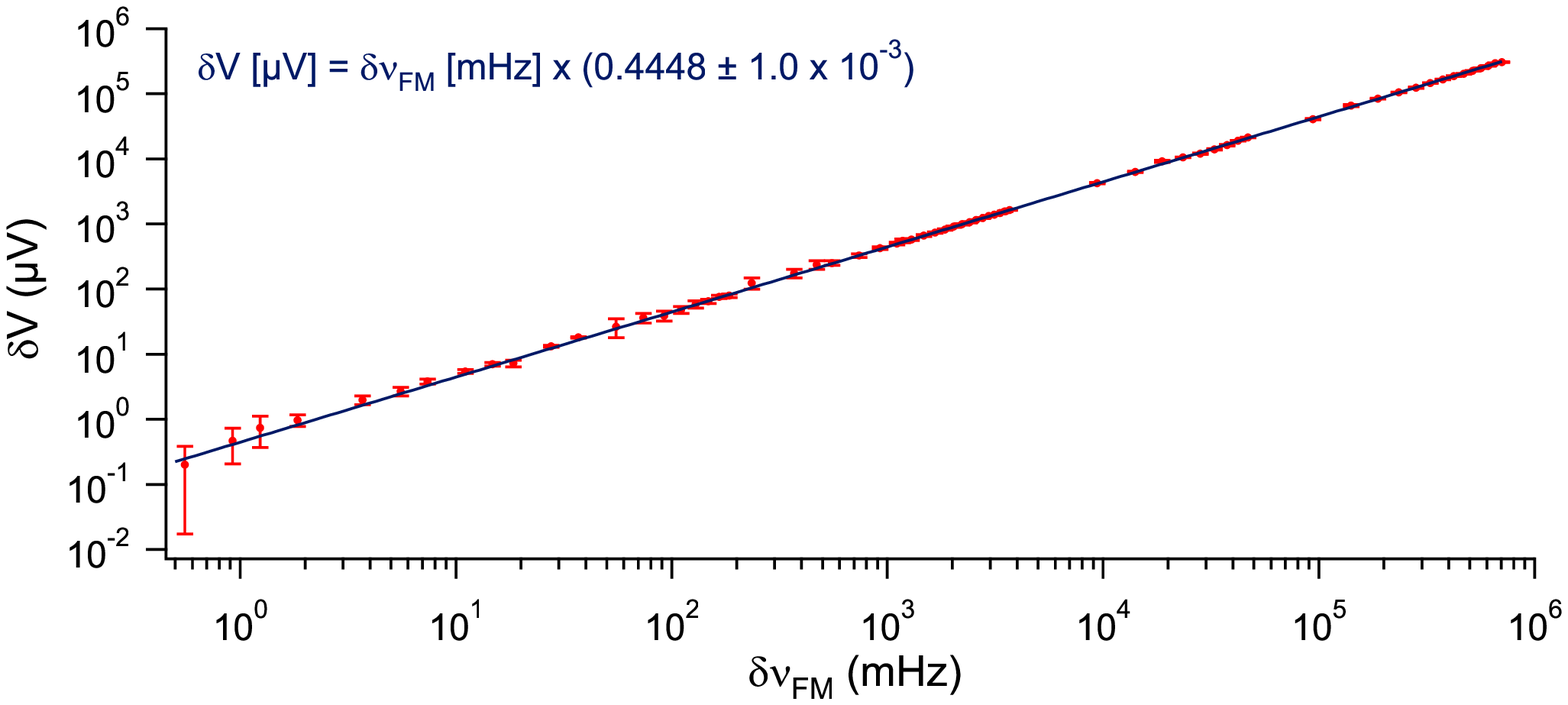}}
\caption{Sensitivity range of our present apparatus. A laser frequency modulation of amplitude $\delta\nu_{\mathrm{FM}}$ at frequency $f_{\mathrm{FM}}$ is created by a sinusoidal voltage applied to an EOM placed on the clockwise beampath. Since this modulation is corrected by the frequency servo-loop, it generates an out-of-phase mismatch of the laser frequency to the cavity resonance on the counterclockwise beam. This mismatch thus appears as a spectral component at $f_{\mathrm{FM}}$ on the counterclockwise error signal. Its amplitude $\delta V$, measured with a lock-in amplifier, is proportional to the frequency excursion $\delta \nu_{\mathrm{FM}}$. The linearity between $\delta V$ and $\delta \nu_{\mathrm{FM}}$ is excellent over more than 6 orders of magnitude, and the sensitivity is 500 $\mu$Hz, achieved with a measurement time of 1000~s.\label{fig:sensitivity}}
\end{figure}

The laser frequency modulation acts as a perturbation and is corrected as such by the servo-loop placed on the cw propagation direction, provided its frequency is within the loop bandwidth. This is definitely the case since the modulation frequency is in the 100-Hz range while the servo-loop bandwidth is in the 100-kHz range. By contrast, the efficient suppression of the frequency modulation on the cw beam creates a discrepancy between the laser and the cavity resonance on the ccw direction, which is detectable by analyzing the ccw error signal at frequency $f_{\mathrm{FM}}$. Using a lock-in amplifier we extract the $f_{\mathrm{FM}}$-spectral component $\delta V$ on the error signal. We present on Fig. \ref{fig:sensitivity} the results of our measurements for different frequency modulation amplitudes and two modulation frequencies  $f_{\mathrm{FM}}$ of 217 and 276~Hz, situated in the region where the noise spectral density is at its lowest (see Fig.~\ref{fig:spectrum}). The experimental points fit nicely on a line over more than 6 orders of magnitude. The frequency sensitivity is 500 $\mu$Hz after a measurement time $\tau \simeq 1000\;$s, which is equivalent to a birefringence sensitivity $\delta n \simeq 2\times 10^{-18}$. Assuming that the noise is white in the region of interest, this yields a noise equivalent birefringence $\gamma_n = \delta n \times \sqrt{4\tau} \simeq 1\times 10^{-16}\;/\sqrt{\mathrm{Hz}}$. 

\section{Outlooks and conclusion \label{conclusion}}

Let us now evaluate how far we are from the ultimate shot-noise limited sensitivity of our present apparatus: for an ideal optical coupling, the mean power impinging onto the reflection photodiode PR$_\mathrm{cw}$ at resonance writes 

$$P_{\mathrm{R\,cw}} = 2P_s + \frac{P_c}{4},$$

\noindent where $P_c$ is the optical power in the carrier and $P_s$ that in each sideband at frequency $\nu \pm \Omega/2\pi$. This generates a statistical white noise with a power spectral density $\gamma_{\mathrm{sn}\,P} = \sqrt{2h\nu P_{\mathrm{R\,cw}}}$, which is converted into frequency noise by the frequency regulation. The frequency noise spectral density $\gamma_{\mathrm{sn}\,\nu}$ is simply obtained by dividing $\gamma_{\mathrm{sn}\,P}$ by the discriminator $D$, which yields

$$\gamma_{\mathrm{sn}\,\nu} = \frac{\sqrt{h\nu}}{2\sqrt 2} \frac{\Delta \nu_{\mathrm{FSR}}}{\mathcal{F}} \sqrt{\frac{P_c/4 + 2P_s}{P_cP_s}}.$$ 

One thus sees that the shot-noise limited sensitivity is inversely proportional to the cavity finesse $\mathcal{F}$ and roughly inversely proportional to the square root of the coupling efficiency, which is represented by $P_c$ in the above equation. For our present values $P_c=10\;$ mW, $P_s=3\;$mW, $\Delta \nu_c \simeq 4\;$kHz, one finally obtains a shot-noise limited sensitivity $\gamma_{\mathrm{sn}\,\nu} \simeq 10\; \mu$Hz$/\sqrt{\mbox{Hz}}$. Converted into birefringence sensitivity, one obtains $\gamma_{\mathrm{sn}\,n} \simeq 1\times 10^{-19}\; /\sqrt{\mbox{Hz}}.$

The sensitivity of our apparatus presently lies 3 orders of magnitude above this value. This estimation is not fully consistent with the spectrum analysis measurements (see Section \ref{stabilization}) which exhibit a noise spectral density only 15 dB above the shot-noise limit. This discrepancy is mainly due to an imperfect optical coupling, 3 to 10 times smaller than the ideal one, and to a finesse of $15000$ instead of $50000$ at the time of our measurements, due to mirror pollution and cavity misalignment. Indeed, both defects contribute to degrade the signal since the discriminator $D$ is proportional to the coupling efficiency and to the finesse. Another noise source originates from the weakness of the voltages corresponding to the smallest frequency modulations we measured, in the mV range, hence very sensitive to environmental noise sources.

The achieved sensitivity is a very encouraging result, and we have already identified several ways to improve it. Let us cite a few improvements that are planned for the near future: first we will optimize the mode matching and cavity alignment, which should allow us to improve our results by at least one order of magnitude. Second, we will place the cavity in a vacuum chamber in order to improve the thermo-acoustic insulation and to decrease the disturbances in the 100-Hz frequency range, where the detection takes place. We also plan to extend the servo-loop bandwidth by a factor 4 to 5 by replacing the Acousto-Optic Modulator by a broadband Electro-Optic Modulator, thus suppressing the delay which severly limits the servo-loop bandwidth. This should in turn increase the servo-loop gain at low frequencies by at least the same factor, provided it is not limited by the operational amplifiers maximum gain. All these developments should allow us to overcome the present state-of-the art by about one order of magnitude \cite{Durand2009} with our present apparatus.

Ultimately, we can reasonably gain 1 to 2 more orders of magnitude on the sensitivity by increasing the cavity finesse and the optical power. Of course, thermal effects on the mirror coatings usually limit the intracavity power. We can ultimately increase it somewhat more by using automatic alignment servos as is done on gravitational waves interferometers \cite{LIGO,VIRGO,GEO600,TAMA,AIGO}.

Now that our apparatus is functional, we will use it to perform some novel measurements. Indeed, it has definitely the required sensitivity to achieve the unprecedented detection of several magneto-electro-optic effects in gases. In addition to the directional anisotropy we have already mentioned, our apparatus also allows us to measure for the first time the Jones magneto-electric birefringence in gases \cite{Rizzo2003,Roth2000}.

\begin{acknowledgments}

The authors thank the LCAR technical staff for support, J.~Vigu\'e for fruitful discussions, M.~Fouch\'e and M.~Reigneau for contributing to a preliminary set-up. G.~B. and C.~R. are grateful to C.~Rizzo for initiating this experiment and for initial support.

\end{acknowledgments}

\end{document}